\documentclass[twocolumn,aps,showpacs,floatfix,superscriptaddress]{revtex4}
\usepackage{amsmath,amssymb,eucal,graphicx,subfigure}

\begin{document}

\title{Dynamics of Microtubule Instabilities}
\author{T. Antal}
\affiliation{Program for Evolutionary Dynamics, Harvard University,
  Cambridge, MA~ 02138, USA}
\author{P. L. Krapivsky}
\affiliation{Center for Polymer Studies and Department of Physics, Boston
  University, Boston, MA~ 02215, USA}
\author{S. Redner}
\affiliation{Center for Polymer Studies and Department of Physics, Boston
  University, Boston, MA~ 02215, USA}

\begin{abstract}

  We investigate an idealized model of microtubule dynamics that involves:
  (i) attachment of guanosine triphosphate (GTP) at rate $\lambda$, (ii)
  conversion of GTP to guanosine diphosphate (GDP) at rate 1, and (iii)
  detachment of GDP at rate $\mu$.  As a function of these rates, a
  microtubule can grow steadily or its length can fluctuate wildly.  For
  $\mu=0$, we find the exact tubule and GTP cap length distributions, and
  power-law length distributions of GTP and GDP islands.  For $\mu=\infty$,
  we argue that the time between catastrophes, where the microtubule shrinks
  to zero length, scales as $e^\lambda$.  We also discuss the nature of the
  phase boundary between a growing and shrinking microtubule.

\end{abstract}

\maketitle

\begin{figure}[ht]
\includegraphics*[width=0.3\textwidth]{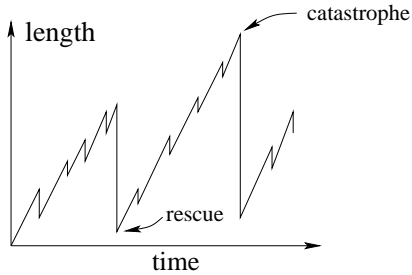}
\caption{Schematic illustration of the growth and catastrophic shrinkage of a
  microtubule as a function of time (adapted from \cite{FBL}).}
\label{l-vs-t}
\end{figure}

Microtubules are linear polymers of the protein tubulin that perform major
organizational tasks in living cells \cite{FBL,VCJ}.  They provide transport
tracks for molecular machines \cite{MK,AH}, and move cellular structures
during cellular processes such as reproduction \cite{VCJ,WMD}.  A surprising
feature of microtubules is that they remain out of equilibrium under fixed
external conditions and can undergo alternating periods of growth and rapid
shrinking (Fig.~\ref{l-vs-t}).

These sudden polymerization changes are driven by the interplay between
growth by the attachment of guanosine triphosphate tubulin complexes (GTP) at
the free end \cite{MK,DM,FBL,DL}, irreversible hydrolysis of GTP into
guanosine diphosphate (GDP) anywhere in the tubule, and subsequent GDP
detachment from the free end of the tubule.  One avenue of theoretical work
on this dynamical instability is based on detailed models of mechanical
stability \cite{JCF,vanburen05}.  For example, a detailed stochastic model of
a microtubule that includes all the thirteen constituent protofilaments has
been investigated in Ref.~\cite{vanburen05}.  By using model parameters that
were inferred from equilibrium statistical physics, VanBuren et al.\
\cite{vanburen05} found characteristics of the tip evolution of a microtubule
that agreed with experimental data \cite{mandelkow91}.

Another approach for modeling the dynamics of microtubules is based on
effective two-state models that describe the dynamics in terms of a switching
between a growing and a shrinking state \cite{DL,Bi,HZ,MKMC,ZLSW,MGGA,hill}.
The essence of many of these models is that a microtubule exists either in a
growing phase (where a GTP cap exists at the end of the microtubule) or a
shrinking phase (without a GTP cap), and that there are stochastic transitions
between these two states.  By tuning parameters appropriately, it is possible
to reproduce the phase changes between the growing and shrinking phases of
microtubules that have been observed experimentally \cite{MK}.  In another
important contribution, Flyvbjerg et al.\ \cite{F} constructed an effective
continuous theory to describe the dynamics of the cap length.

What is missing in these models is a clear connection between microscopic
parameters and the evolution of the spatial structure of the microtubule.
Most of the models discussed thus far are too complicated to permit a
complete solution.  Thus the present work is motivated by the goal of
formulating and solving a minimal and idealized coarse-grained model of
microtubule dynamics that incorporates the main features of growth and
shrinking.  We view this idealized model as an Ising-like description that
captures the most interesting features of microtubule dynamics even though
the connection between model parameters and experimental variables is
indirect.  One of the main advantages of the model is its simplicity so that
many of its geometrical and dynamical features can determined analytically.
These solutions reveal many rich geometrical and time-dependent features that
should help our understanding of real microtubule dynamics.

The model studied in this paper \cite{model} treats a microtubule as a linear
polymer that consists of GTP or GDP monomers that we denote as $+$ and $-$,
respectively.  To emphasize this connection between chemistry and the model,
we write the former as GTP$^+$ and the latter as GDP$^-$.  The state of a
microtubule evolves by the following steps:

1. Attachment of GTP$^+$ at the end of a microtubule:
\begin{eqnarray*}
    |\cdots \rangle \Longrightarrow |\cdots +\rangle\qquad {\rm rate}~ \lambda
\end{eqnarray*}

2. Independent conversion of each GTP$^+$ to GDP$^-$:
\begin{eqnarray*}
 ~~~~~~~~~~~|\cdots +\cdots\rangle \Longrightarrow |\cdots-\cdots \rangle \qquad {\rm rate}~ 1.
\end{eqnarray*}

3. Detachment of a GDP$^-$ from the microtubule end:
\begin{eqnarray*}
|\cdots-\,\rangle\Longrightarrow |\cdots\rangle\qquad {\rm rate}~ \mu.
\end{eqnarray*}
Here the symbols $|$ and $\rangle$ denote the terminal and the free end of
the microtubule.  It is worth mentioning that these steps are similar to
those in a recently-introduced model of DNA sequence evolution \cite{MLA},
and that some of the results about the structure of DNA sequences seem to be
related to our results about island size distributions in microtubules.

With the above steps for the evolution of a microtubule, we find that the
$(\lambda,\mu)$ phase plane separates into a region where the microtubule
grows, on average, and a phase where the microtubule length is finite.  On
the boundary between these two regions, the microtubule length fluctuates
wildly.  To understand these different phases, we first focus on the extreme
cases of no detachment $\mu=0$ and infinite detachment rate $\mu=\infty$,
where we can analytically solve the microtubule structure.

{\tt No Detachment:} Here GTP$^+$ monomers attach at rate $\lambda$ and
convert to GDP$^-$ at rate 1.  The probability $\Pi_N(t)$ that the
microtubule consists of $N$ GTP$^+$ monomers at time $t$ evolves according to
\begin{equation}
\label{PNt}
\frac{d\Pi_N}{dt} = -(N+\lambda)\Pi_N+\lambda\Pi_{N-1}+(N+1)\Pi_{N+1}.
\end{equation}
The loss term $(N+\lambda)\Pi_N$ accounts for the attachment of a GTP$^+$ at
the end of the microtubule of length $N$ with rate $\lambda$ and the
conversion events GTP$^+$ $\to$ GDP$^-$ that occur with total rate $N$.  The
gain terms can be explained similarly.  This equation can be solved by the
generating function method and the final result is
\begin{equation}
\label{Poisson-t}
\Pi_N(t) =  \frac{[\lambda(1-e^{-t})]^N}{N!}\,e^{-\lambda(1-e^{-t})}.
\end{equation}
From this Poisson distribution, the mean number of GTP$^+$ monomers and its
variance are
\begin{equation}
\label{N-av-var}
\langle N\rangle = \langle N^2\rangle - \langle N\rangle^2 = \lambda(1-e^{-t}).
\end{equation}

Similarly, the length distribution $P(L,t)$ of the microtubule evolves
according to the master equation
\begin{equation}
\label{PLt}
\frac{d P(L,t)}{d t} = \lambda\left[P(L-1,t)-P(L,t)\right]
\end{equation}
For the initial condition $P(L,0)=\delta_{L,0}$, the solution is again the
Poisson distribution
\begin{equation}
\label{PLt-sol}
P(L,t) = \frac{(\lambda t)^L}{L!}\,e^{-\lambda t}
\end{equation}
{}from which the growth rate of
the microtubule and the diffusion coefficient of the tip are
$V=\lambda$ and $D=\lambda/2$.

\begin{figure}[ht]
\includegraphics*[width=0.4\textwidth]{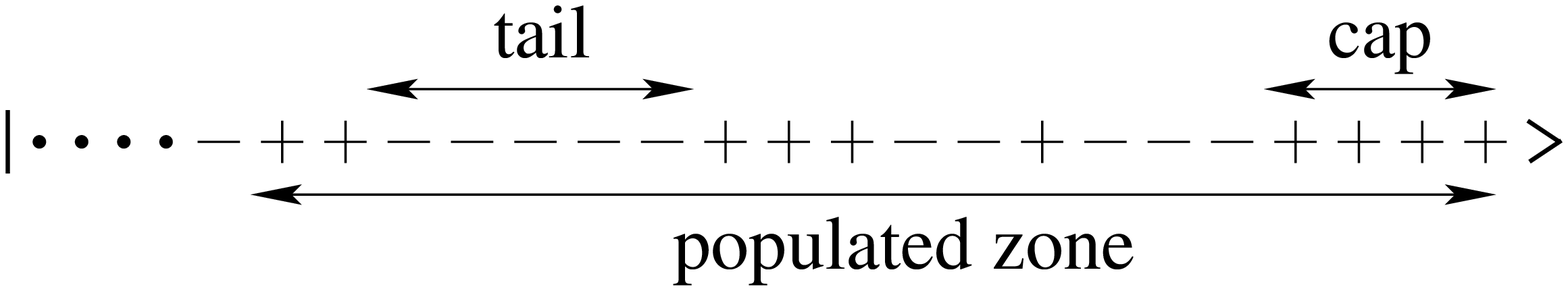}
\caption{Representative configuration of a microtubule, with a cap of length
  4, three GTP$^+$ islands of lengths 1, 3, and 2, and three GDP$^-$ islands of
  lengths 3, 2, and a ``tail'' of length 5.  The rest of the microtubule
  consists of GDP$^-$.}
\label{cap-fig}
\end{figure}

Because of the conversion of GTP$^+$ to GDP$^-$, the tip of the microtubule
is comprised predominantly of GTP$^+$, while the end exclusively consists of
GDP$^-$.  The region from the tip until the first GDP$^-$ is known as the
{\em cap} (Fig.~\ref{cap-fig}) and it plays a fundamental role in microtubule
function.  We use a master equation approach to determine the cap length
distribution \cite{all}.

Consider a cap of length $k$.  Its length increases by 1 due to the
attachment of a GTP$^+$ at rate $\lambda$.  The conversion of any GTP$^+$
into a GDP$^-$ at rate 1 reduces the cap length from $k$ to an arbitrary
value $s<k$.  These processes lead to the following master equation for the
probability $n_k$ that the cap length equals $k$:
\begin{equation}
\label{nk}
\dot n_k = \lambda(n_{k-1}-n_k)-kn_k+\sum_{s\geq k+1}n_s.
\end{equation}
Equation~\eqref{nk} remains valid for $k=0$ if we set $n_{-1}\equiv 0$.  We
solve for the stationary distribution by summing the first $k-1$ of
Eqs.~\eqref{nk} with $\dot n_k$ set to zero to obtain
\begin{equation}
 n_{k-1} = \frac{k}{\lambda} \sum_{s\geq k}n_s.
\end{equation}
The cumulative distribution, $ N_k=\sum_{s\geq k}n_s$, thus satisfies the
recursion $N_k = \lambda N_{k-1}/(k+\lambda)$.  Using the normalization
$N_0=1$ and iterating, we obtain the solution in terms of the Gamma function
\cite{AS}:
\begin{equation}
\label{Nk-soln}
 N_k = \frac{\lambda^k\, \Gamma(1+\lambda)}{\Gamma(k+1+\lambda)}.
\end{equation}
Hence the cap length distribution is
\begin{equation}
\label{nk-sol}
 n_k = \frac{\Gamma(1\!+\!\lambda)}{\Gamma(k\!+\!2\!+\!\lambda)}\, (k+1)\lambda^k
\end{equation}

From this distribution, we find that in the realistic limit of $\lambda\gg 1$
the average cap length is given by 
\begin{equation}
\label{kav-large}
\langle k\rangle\to \sqrt{\frac{\pi\lambda}{2}}\quad {\rm as}\quad \lambda\to\infty
\end{equation}
Thus even though the average number of GTP$^+$ monomers equals $\lambda$,
only $\sqrt{\lambda}$ of them organize themselves into the microtubule cap.

At a finer level of resolution, we determine the distribution of island sizes
in the GTP$^+$-populated zone of a microtubule (Fig.~\ref{cap-fig}).  When
$\lambda\to\infty$, both the length of the cap and the length of the region
that contains GTP$^+$ become large and a continuum description is
appropriate.  Since the residence time of each monomer increases linearly
with distance from the tip and the conversion GTP$^+$ $\to$ GDP$^-$ occurs
independently, the probability that a monomer remains a GTP$^+$ decays
exponentially with distance from the tip.  This fact alone is sufficient to
derive all the island distributions.

Consider first the length $\ell$ of the populated zone (Fig.~\ref{cap-fig}).
For a monomer at a distance $x$ from the tip, its residence time is
$\tau=x/\lambda$ for large $\lambda$.  Thus the probability that this monomer
remains a GTP$^+$ is $e^{-\tau}=e^{-x/\lambda}$.  We thus estimate $\ell$
from the extremal criterion \cite{extremal}
\begin{equation}
\label{ell}
1=\sum_{x\geq \ell} e^{-x/\lambda}=(1-e^{-1/\lambda})^{-1}e^{-\ell/\lambda},
\end{equation}
that merely states that there is of the order of a single GTP$^+$ further
than a distance $\ell$ from the tip.  When $\lambda$ is large,
$(1-e^{-1/\lambda})^{-1}\to \lambda$, and the length of the active region
scales as
\begin{equation}
\label{ell-sol}
\ell=\lambda\ln\lambda
\end{equation}

The probability that the cap has length $k$ is given by
\begin{equation*}
(1-e^{-(k+1)/\lambda})\prod_{j=1}^k e^{-j/\lambda}.
\end{equation*}
The product ensures that all monomers between the tip and a distance $k$ from
the tip are GTP$^+$, while the prefactor gives the probability that a monomer a
distance $k+1$ from the tip is a GDP$^-$.  Expanding the prefactor for large
$\lambda$ and rewriting the product as the sum in the exponent, we obtain
\begin{equation}
\label{nk-cont}
n_k\sim \frac{k+1}{\lambda}\,\, e^{-k(k+1)/2\lambda},
\end{equation}
a result that agrees with the large-$\lambda$ limit of the exact result for
$n_k$ in Eq.~\eqref{nk-sol}.

Similarly, the probability to find a positive GTP$^+$ island of length $k$
that occupies sites $x+1,x+2,\ldots,x+k$ is
\begin{equation}
\label{isl-x}
(1-e^{-x/\lambda})(1-e^{-(x+k+1)/\lambda})\prod_{j=1}^k e^{-(x+j)/\lambda}.
\end{equation}
The two prefactors ensure that sites $x$ and $x+k+1$ are GDP$^-$, while
the product ensures that all sites between $x+1$ and $x+k$ are GTP$^+$.

Most islands are far from the tip and they are short, $k\ll x$, so that
\eqref{isl-x} simplifies to $(1-e^{-x/\lambda})^2 e^{-kx/\lambda}$.  The
total number of GTP$^+$ islands of length $k$ is obtained by summing this
island density over all $x$.  Since $\lambda\gg 1$, we replace the summation
by integration and obtain
\begin{equation}
\label{isl-k}
I_k\!=\!\int_0^\infty \!\!dx\,(1-e^{-x/\lambda})^2 e^{-kx/\lambda}=
\frac{2\lambda}{k(k\!+\!1)(k\!+\!2)}.
\end{equation}

By similar reasoning, we find that the density of negative GDP$^-$ islands of
length $k\ll x$ with one end at $x$ is given by $e^{-2x/\lambda}
(1-e^{-x/\lambda})^k$.  The total number of negative islands of length $k$ is
therefore
\begin{equation}
\label{isl-k-negative}
J_k=\int_0^\infty dx\,e^{-2x/\lambda} (1-e^{-x/\lambda})^k =\frac{\lambda}{(k+1)(k+2)}.
\end{equation}
Again, we find a power-law tail for the GDP$^-$ island size distribution, but
one that is much broader than the corresponding GTP$^+$ distribution.
Strikingly, these two distributions are of the same form as those found for
the degree distribution of growing networks \cite{network1,network2}.  The
total number of GDP$^-$ monomers within the populated zone is $\sum_{k\geq
  1}kJ_k$.  While this sum formally diverges, we invoke the upper size
cutoff, $k_*\sim \lambda$ that again follows from an extremal criterion
\cite{extremal} to obtain $\sum_{k\geq 1}kJ_k\simeq \lambda\ln\lambda$.
Since the length of the populated zone $\ell\sim \lambda\ln \lambda$, we thus
see that this zone predominantly consists of GDP$^-$ islands.

\begin{figure}[ht]
\includegraphics*[width=0.4\textwidth]{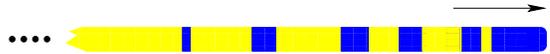}
\caption{Cartoon of a microtubule.  The GTP$^+$ regions (dark) get shorter
  further from the tip that advances as $\lambda t$, while the GDP$^-$
  regions (light) get longer.}
\label{cartoon}
\end{figure}

In analogy with the cap, consider now the ``tail''---the last island of
GDP$^-$ within the populated zone (Fig.~\ref{cap-fig}).  The probability
$m_k$ that it has length $k$ is
\begin{equation}
\label{last}
m_k = e^{-\ell/\lambda} (1-e^{-\ell/\lambda})^k.
\end{equation}
Hence, by summing the geometric series and using $\ell=\lambda\ln \lambda$
from \eqref{ell-sol}, the average length of the tail is
\begin{equation}
\label{last-length}
\langle k\rangle \equiv \sum_{k\geq 1} km_k = e^{\ell/\lambda}-1\to \lambda.
\end{equation}
Thus the tail is (on average) much longer than the cap.  In summary, the
microtubule consists of GTP$^+$ islands that systematically get shorter away
from the tip and vice versa for GDP$^-$ islands (Fig.~\ref{cartoon}).

{\tt Instantaneous Detachment:} For $\mu>0$, a microtubule can recede if the
monomer(s) at its tip are GDP$^-$.  If there are many such GDP$^-$ end
monomers the microtubule length can fluctuate wildly under steady conditions.
Here we focus on the limiting case of instantaneous detachment, $\mu=\infty$.
As soon as the monomer at the tip changes from a GTP$^+$ to a GDP$^-$, this
monomer and any contiguous GDP$^-$ monomers detach immediately.  We term such
an event an {\em avalanche of size $k$}.  If the avalanche encompasses the
entire microtubule, we have a {\em global catastrophe}.  The probability for
such a catastrophe to occur is
\begin{equation}
\label{sweep}
\mathcal{C}(\lambda)=\frac{1}{1+\lambda}\prod_{n=1}^\infty (1-e^{-n/\lambda}).
\end{equation}
The factor $(1+\lambda)^{-1}$ gives the probability that the monomer at the
tip converts to a GDP$^-$ before the next attachment event, while the product
gives the probability that all other monomers in the microtubule are GDP$^-$.  In
principle, the upper limit in the product is set by the microtubule length.
However, for $n>\lambda$, each factor in the product is close to 1 and the
error made in extending the product to infinity is small. 

The asymptotic behavior of the infinite product in Eq.~\eqref{sweep} is found
by first expressing it in terms of the Dedekind $\eta$ function \cite{apostol}
\begin{equation}
\label{eta}
\eta(z)=e^{i\pi z/12}\prod_{n=1}^\infty (1-e^{2\pi i nz}),
\end{equation}
and then recalling a remarkable identity satisfied by this function,
$\eta(-1/z)=\sqrt{-i z}\,\eta(z)$.  Using these facts, the probability of a
catastrophe is given by
\begin{eqnarray}
\label{S}
\mathcal{C}(\lambda)&=&\frac{\sqrt{2\pi\lambda}}{1+\lambda} \,\,e^{-\pi^2\lambda/6}\, e^{1/24\lambda}\,
\prod_{n\geq 1} (1-e^{-4\pi^2\lambda  n})\nonumber \\
&\sim& \sqrt{\frac{2\pi}{\lambda}}\,\,e^{-\pi^2\lambda/6}.
\end{eqnarray}
Since the time between catastrophes scales as the inverse of the occurrence
probability, this inter-event time becomes astronomically long for large
$\lambda$.

\begin{figure}[ht]
\includegraphics*[width=0.175\textwidth]{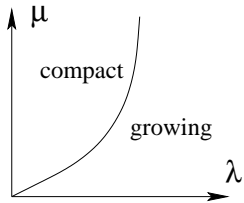}
\caption{Schematic phase diagram for a microtubule in the $\lambda$-$\mu$
  parameter space.}
\label{phase-diag}
\end{figure}

Similar arguments can be used to calculate the size distribution of finite
avalanches.  We first notice that the configuration $|+\underbrace{-\cdots
  -}_{k-1} +\rangle$ occurs with probability
\begin{equation*}
\prod_{n=1}^{k-1} (1-e^{-n/\lambda}).
\end{equation*}
Then multiplying by the probability that the monomer at the tip converts
before the next attachment event gives the probability for an avalanche of
size $k$:
\begin{equation}
\label{Ak-gen}
A_k = (1+\lambda)^{-1}\prod_{n=1}^{k-1} (1-e^{-n/\lambda})
\end{equation}
Using $1-e^{-n/\lambda}\approx n/\lambda-n^2/(2\lambda^2)$, Eq.~\eqref{Ak-gen}
becomes
\begin{equation*}
A_k = \lambda^{-k}\Gamma(k)\prod_{n=1}^{k-1} \left(1-\frac{n}{2\lambda}\right)\sim
\lambda^{-k}\Gamma(k)\,e^{-k^2/4\lambda}.
\end{equation*}

{\tt General Growth Conditions:} For arbitrary attachment and detachment
rates $\lambda$ and $\mu$, we can approximately map out the boundary between
different phases of microtubule dynamics in the $\lambda,\mu$ parameter
space.  For $\lambda, \mu \ll 1$, the unit conversion rate GTP$^+\to$ GDP$^-$
is much faster than the rates $\lambda, \mu$ of the other dynamical
microtubule processes.  Hence we assume that conversion is instantaneous.
Consequently, a microtubule consists of a string of GDP$^-$ monomers $|\cdots
---\rangle$ in which the tip advances at rate $\lambda$ and retreats at rate
$\mu$.  Thus the tip performs a biased random walk, and its velocity is
$V(\lambda, \mu) = \lambda-\mu$ when $\lambda>\mu$.  The boundary between the
growing phase, where the average tubule length grows as $Vt$, and the compact
phase, where the average tubule length is finite, is found by setting $V=0$.
This condition gives $\mu_* = \lambda$ when $\lambda\ll 1$.  On this phase
boundary, the average tubule length grows as $\sqrt{t}$.

For the phase boundary for large $\mu$, consider first $\mu=\infty$.  Since
the leading behavior for the probability of a catastrophe scales as
$e^{-\pi^2\lambda/6}$, the typical time between catastrophes is
$e^{\pi^2\lambda/6}$.  Since $V\approx \lambda$ for large $\lambda$, the
typical length of a microtubule just before a catastrophe is again (ignoring
all power-law factors) $e^{\pi^2\lambda/6}$.  Suppose now the detachment rate
$\mu$ is very large but finite.  The microtubule is compact if the time to
shrink a microtubule of length $e^{\pi^2\lambda/6}/\mu$ is smaller than the
time $\lambda^{-1}$ required to generate a GTP$^+$ by $|\cdots-\rangle
\Longrightarrow |\cdots-+\rangle$ and thereby stop the shrinking.  We
estimate the location of the phase boundary by equating these two times to
give $\mu_*\sim e^{\pi^2\lambda/6}$ when $\lambda\gg 1$
(Fig.~\ref{phase-diag}).

To summarize, our model predicts rich microtubule dynamics as a function of
GTP$^+$ attachment and GDP$^-$ detachment.  In the growing phase, GTP$^+$ and
GPD$^-$ organize into alternating domains, with gradually longer GTP$^+$
domains and gradually shorter GDP$^-$ domains toward the tip of the
microtubule. The size distributions of these two species are exact power laws
with respective exponents of $3$ and $2$.  Between the limiting cases of a
finite-length and a growing microtubule, its length can fluctuate wildly
under steady external conditions.  This unusual behavior emerges naturally in
our model.  From a simple probabilistic approach and in the limit
instantaneous detachment of GDP$^-$ ($\mu=\infty$), we found that the time
between catastrophes, where the microtubule shrinks to zero length, scales
exponentially with the attachment rate $\lambda$.  Thus for large $\lambda$,
the microtubule will grow essentially freely for a very long time before
undergoing a catastrophe.

Acknowledgments: We thank Bulbul Chakraborty for introducing us to this
problem and the model presented here.  We also acknowledge financial support
to the Program for Evolutionary Dynamics at Harvard University by Jeffrey
Epstein and NIH grant R01GM078986 (TA), NSF grant CHE0532969 (PLK), and NSF
grant DMR0535503 (SR) at Boston University.

\end{document}